\documentclass{cup-hpl}
\usepackage[none]{hyphenat}
\usepackage{lipsum}
\usepackage{makecell}
\usepackage{microtype}

\begin{document}


\shorttitle{Residual pump diagnostic thin-disk laser}
\shortauthor{H. Jo et al}

\title{Residual pump diagnostics of lasing-state absorption in multipass pumped Yb:YAG thin-disk lasers}

\author[1,*]{Hanjin Jo\corresp{H. Jo, HiLase Centre, 252 41 Dolní Břežany, Czech Republic. \email{hanjin.jo@hilase.cz}}}
\author[1]{Jiří Mužík}
\author[1]{Pawel Sikocinski}
\author[1]{Michal Chyla}
\author[1]{Yoann Levy}
\author[1]{Martin Smrž}
\author[1]{Tomáš Mocek}

\address[1]{HiLASE Centre, Institute of Physics of the Czech Academy of Sciences, Za Radnicí 828, 252 41 Dolní Břežany, Czech Republic}

\begin{abstract}
Residual pump power was used as a spatially integrated diagnostic of inversion-dependent absorption in 32-pass pumped Yb:YAG thin-disk lasers. A quasi-three-level rate-equation model was coupled to accumulated pump flux, intracavity signal flux, and an effective ASE-induced depletion term. Two disks were tested under non-lasing and multimode lasing conditions. The analysis reproduced residual pump fractions within 2.4\% in non-lasing and 1.8\% in multimode operation, and output power within 1.3\% relative error. Stimulated emission reduced the steady-state inversion and increased the effective pump-band absorption relative to the non-lasing case, while the absorption remained below the room-temperature small-signal value. The same absorption state was then used to estimate volumetric heat-load trends and to assess pump-pass number and output-coupler transmission for high-power thin-disk scaling.

\end{abstract}


\maketitle

\section{Introduction}
Thin-disk (TD) gain medium is attractive for high-power operation due to efficient heat removal. In high-power operation, however, reliable design requires not only efficient heat extraction but also an accurate estimate of the absorbed pump power and the local heat load. This requirement is particularly important in multi-pass pumped Yb:YAG thin-disks, where the pump absorption experienced by each pass is coupled to stimulated emission, amplified spontaneous emission, and heat load of TD. In high-power thin-disk lasers, thermal deformation and thermo-optic phase distortion degrade resonator-mode quality and can reduce the extractable output power\cite{LHeat1, LHeat2, LHeat3, LHeat4,L_Cryo1,TDL_diff1}. Therefore, pump-module design requires an accurate estimate of the absorbed pump power and the associated heat load under lasing operation, rather than the small-signal absorption coefficient alone.

Practical thin-disk laser design often starts from output power, slope efficiency, resonator loss, and output-coupler transmission\cite{TDL_Model1, TDL_Model2, TDL_Model3}. An output-oriented approach is necessary, but it does not uniquely determine the pump absorption occurring inside a multi-pass pump module. In a quasi-three- level medium, the inversion modifies the net pump absorption, while stimulated emission, reabsorption, and ASE modify the inversion itself. Such coupling between population inversion, absorption, and heat load is present in general gain media, but it becomes more pronounced in quasi-three-level systems, where reabsorption and net gain depend strongly on temperature and inversion\cite{LV3}. Consequently, residual pump power, absorption, stimulated emission at the signal wavelength and heat load are coupled through the same population dynamics and should be analyzed together.

In quasi-three-level Yb:YAG, pump absorption is modified by the instantaneous population inversion, while stimulated emission, reabsorption, and ASE modify the inversion\cite{Peterson2011,ASE1,ASE2, ASE3}. In a multi-pass pump module, the residual pump power after the final pass therefore provides a spatially integrated observable of the lasing-state absorption. Thus, the residual pump after the final pass carries integrated information on the absorption under lasing operation and can be used as an experimentally accessible diagnostic of the absorption state.

In this work, the final residual pump power is treated as a spatially integrated diagnostic observable for the absorption state of a multipass pumped Yb:YAG thin-disk laser. The residual pump is not used only as an unabsorbed-power loss term. Instead, it is used as an absorption-side constraint on the gain medium under non-lasing (NL) and multimode lasing (MM) operation. A quasi-three-level rate-equation model is utilized as an analyzing tool for the measured residual pump, the inversion-dependent pump-band absorption, resonator-mode extraction, and ASE-induced inversion depletion. By validating the residual pump fraction and the output power for two Yb:YAG thin disks, we show that residual-pump diagnostics can distinguish small-signal, non-lasing, and lasing-state absorption. The evaluated absorption state is then used to estimate heat-load trends and to examine pump-pass number, output-coupler transmission, and ASE-management effects relevant to high-power thin-disk laser scaling.

\section{Residual-pump diagnostic principle}

The measured quantity used in this work is the final residual pump power after the multipass pump sequence. In a 32-pass thin-disk pump module, residual pump power contains the accumulated effect of pump depletion over all disk traversals. After accounting for pump-module throughput, the residual pump fraction provides a pass-integrated constraint on the pump absorption. The rate-equation calculation is then used to determine the steady-state inversion, resonator-mode extraction, and ASE-induced inversion depletion that are consistent with the measured residual pump fraction.

In Yb:YAG gain medium, pump absorption creates a population inversion in the upper laser manifold. The inversion is depleted by stimulated emission and by radiative loss channels such as spontaneous emission and ASE\cite{Peterson2011,ASE1}. With negligible energy transfer upconversion, excited-state absorption, and fast intra-manifold non-radiative relaxation, any absorbed pump power not extracted as signal ultimately appears as heat, motivating a quasi‑three‑level rate‑equation description~\cite{YbArt1,YbArt2}.

\begin{figure*}[t]
\centering\includegraphics[width=\textwidth]{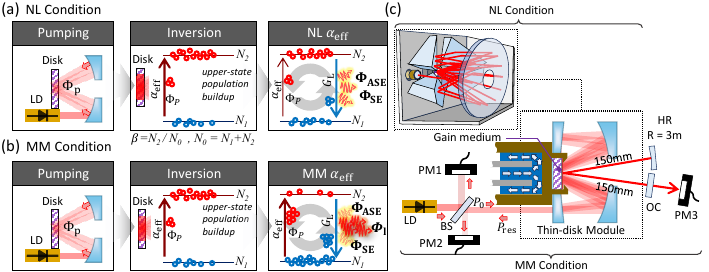}
\caption{\label{fig1} Residual-pump diagnostic concept for a multipass-pumped Yb:YAG thin-disk laser. Block-diagram schematics of the quasi-three-level rate-equation model for (a) the non-lasing (NL) case, including pump absorption with spontaneous emission (SE) and amplified spontaneous emission (ASE) only, and (b) the multimode lasing (MM) case, additionally including signal-induced stimulated emission. (c) Experimental setup for measuring the residual pump power. Abbreviations are following; LD: laser diode, BS: beam splitter, PM: powermeter, HR: high-reflectivity mirror and OC: output-coupler.} 
\end{figure*}

To use the residual pump as a diagnostic of the absorption state in a multi-pass pumped thin disk, the theoretical approach is organized into four coupled steps: (i) calculation of the population inversion, (ii) evaluation of the inversion- dependent pump absorption including ASE-induced depletion, (iii) calculation of the intracavity signal photon flux synchronized with the multi-pass pump sequence, and (iv) estimation of the volumetric heat load. These quantities are updated iteratively until a steady-state solution is obtained. This framework allows the NL and MM condition to be treated within the same model: in NL, the resonator-mode photon flux is absent and only ASE contributes at the signal wavelength, whereas in MM, the intracavity signal photon flux must be included to describe stimulated emission, gain saturation, and gain clamping, as illustrated in figure \ref{fig1}(a) and (b).

\subsection{Definition of photon flux and rate equation}

The TD gain medium is treated as uniform over the disk thickness and uses volume-averaged populations. Consider a thin-disk Yb:YAG gain medium of thickness $L_\mathrm{d}$ and active-ion density $N_0$, pumped in a multi-pass geometry. Let $N_2(t)$ and $N_1(t)$ denote the upper- and lower-laser-level population densities, respectively, with $N_0 = N_1 + N_2$. For the population inversion ratio $\beta = N_2/N_0$, then

\begin{align}\label{RateEq}
\begin{split}
\dfrac{d\beta}{dt} = -\gamma_{12}\beta+(\sigma_\mathrm{a}^\mathrm{p}\cdot (1-\beta)-\sigma_\mathrm{e}^\mathrm{p}\cdot \beta)\Phi_\mathrm{p}(t_k) \\+ (\sigma_\mathrm{a}^\mathrm{l}\cdot (1-\beta)-\sigma_\mathrm{e}^\mathrm{l}\cdot \beta)\Phi_\mathrm{l}(t_k)
\end{split}
\end{align}

The theoretical approach begins with synchronizing the timing of the multi-pass pump with that of stimulated emission. Time is discretized as $t_k = t_0 + k\Delta t$ with $t_0$ as onset time of the pump and $\Delta t = L_\mathrm{c}/c$. We introduce $L_\mathrm{c}$ for temporal consistency of absorption and emission, where $L_\mathrm{c}$ is the round-trip propagation distance of the beam reflected from the TD in the thin-disk Module, which contains parabolic mirror and deflection prism. This approach allows us to simultaneously compute absorption and emission for population inversion in the time domain. 

The discretized time coordinate is used as a numerical coordinate for pump-pass-resolved absorption and resonator-mode build-up. It is not interpreted as a directly measured transient waveform. Only the converged steady-state values of residual pump, output power, effective absorption, inversion ratio, and heat load are compared with experiment or used for design interpretation.

The radiative decay rate is $\gamma_{12} = 1/\tau_f$, where $\tau_f$ is the fluorescence lifetime, and $\sigma_{a,e}^\mathrm{p,l}$ denotes the temperature dependent absorption/emission cross sections at the pump $(\mathrm{p})$ and signal $(\mathrm{l})$ wavelengths. We define the photon flux ($\boldsymbol{\Phi}_\mathrm{p,l}$ photon rate per unit area) with the area $A_\mathrm{p,l}$ and the energy $E_\mathrm{p,l}$ of the pump and signal as

\begin{align}\label{PhotonFlux}
\begin{split}
\boldsymbol{\Phi}_\mathrm{p,l} = \dfrac{I_\mathrm{p,l}}{h\cdot \nu_\mathrm{p,l}}=\dfrac{dE_\mathrm{p,l}}{dtA_\mathrm{p,l}}\dfrac{1}{h\cdot \nu_\mathrm{p,l}}
\end{split}
\end{align}

The pump ($A_\mathrm{p}$) and signal ($A_\mathrm{l}$) mode areas are assumed to share the same effective $1/\mathrm{e^2}$ intensity area. The equality of $A_\mathrm{p}=A_\mathrm{l}=A\equiv\pi R_\mathrm{p}^2$ is an effective multimode overlap approximation. Under steady-state conditions  $(d\beta/dt = 0)$, the initial population inversion ratio $(\beta(t_0))$ becomes

\begin{align}\label{RateSS}
\begin{split}
\beta(t_0) =\cfrac{N_2}{N_0} \\= \dfrac{\sigma_\mathrm{a}^\mathrm{(p)}\cdot\Phi_\mathrm{p}(t_0) + \sigma_\mathrm{a}^\mathrm{(l)}\cdot\Phi_\mathrm{l}(t_0)}{\gamma_{12}+(\sigma_\mathrm{a}^\mathrm{(p)}+\sigma_\mathrm{e}^\mathrm{(p)})\cdot\Phi_\mathrm{p}(t_0) + (\sigma_\mathrm{a}^\mathrm{(l)}+\sigma_\mathrm{e}^\mathrm{(l)})\cdot\Phi_\mathrm{l}(t_0)}
\end{split}
\end{align}

In the NL condition, the signal photon flux only contains ASE photon flux $\boldsymbol{\Phi}_\mathrm{l}(t_k) = \boldsymbol{\Phi}_\mathrm{ASE}$. On the other hand in MM, the signal photon flux on the disk should include stimulated emission which is described in section $\ref{ICSig}$. The population inversion should be updated in the transient response with mutual interaction between emission and absorption. The signal intensity evolution is modeled as $dI_\mathrm{l}/dz = g_kI_\mathrm{l}$ from small signal gain coefficient $g_k$ where the gain coefficient in the time domain is

\begin{align}\label{SSGain}
\begin{split}
g(t_k)\equiv g_k = N_0(\sigma_\mathrm{e}^\mathrm{(l)}\beta(t_k)-\sigma_\mathrm{a}^\mathrm{(l)}(1-\beta(t_k))) \\ G_k = \exp{(g_k\cdot L_\mathrm{d})}
\end{split}
\end{align}
The single-pass gain ($G_k$) is calculated through the disk thickness ($L_\mathrm{d}$). At the pump wavelength, the inversion dependent effective absorption coefficient is

\begin{align}\label{AbsEff}
\begin{split}
\alpha_\mathrm{eff}(t_k) = N_0(\sigma_\mathrm{a}^\mathrm{(p)}(1-\beta(t_k)) - \sigma_\mathrm{e}^\mathrm{(p)}\beta(t_k))
\end{split}
\end{align}

\subsection{Absorption and ASE}

Assuming a uniform $\alpha_\mathrm{eff}(t_k)$ across the disk thickness , the transmitted pump power after one traversal ($P^{\mathrm{T}}_i(t_k)$) is updated as

\begin{align}\label{PTrans}
\begin{split}
P^{\mathrm{T}}_i(t_k) = P_{i-1}^{\mathrm{T}}(t_{k-1})\exp{(-\alpha_\mathrm{eff}(t_k) L_\mathrm{d}/\cos{\theta_{inc}})}R_\mathrm{PM},\\ ( P_{0}^{\mathrm{T}}(t_k) = P_0 )
\end{split}
\end{align}

where $i$ denotes the number of pump-pass, $\theta_{inc}$ is incidence angle of pump, $P_0$ is incident pump power and $R_\mathrm{PM}$ is the reflectivity of parabolic mirror and prism in the thin-disk module. It is necessary to calculate the effective pump photon flux, because the pump-passes are overlapped in the same spatial region. With the number of pump-passes ($M$), and pump beam propagating through the same pumped region, the effective pump photon flux $(\boldsymbol{\Phi}_\mathrm{p}(t_k))$ used in the rate equation is computed from the sum of the transmitted power in the local area $P_\mathrm{T,OL}$

\begin{align}\label{PhFlux_Pump}
\begin{split}
\boldsymbol{\Phi}_\mathrm{p}(t_k)=\dfrac{P_\mathrm{T,OL}}{A_\mathrm{p}}\dfrac{1}{h\nu_\mathrm{p}}=\dfrac{\sum_{i=0}^{M-1} P_i^\mathrm{T}(t_k)}{A_\mathrm{p}}\dfrac{1}{h\nu_\mathrm{p}}
\end{split}
\end{align}

After each pump-pass, the absorption is updated according to the instantaneous inversion, and the spatially overlapped pump photon flux ($\boldsymbol{\Phi}_\mathrm{p}(t_k)$) is then calculated using the updated absorption. The residual pump power $(P_\mathrm{res})$ then can be evaluated as

\begin{align}\label{ResPump}
\begin{split}
P_\mathrm{res} = P^\mathrm{T}_{M}(t_k)
\end{split}
\end{align}

It highlights that the $P_\mathrm{res}$ is used as diagnostic observable for pump absorption state. The population inversion is driven by the total pump photon flux incident on the pumped volume, whereas the diagnostic measurement is the sequentially depleted pump power after the final pass. Therefore, $P_\mathrm{res}$ sets a limit on the steady-state average absorption across the whole multi-pass. The residual-pump diagnostic used here is power-based and spatially integrated, and it does not reconstruct the residual beam profile.

The ASE is included as an additional inversion-depletion channel since the ASE photon flux $\boldsymbol{\Phi}_\mathrm{ASE}$ contributes to stimulated emission. The ASE term is introduced as an effective inversion-depletion channel rather than as a full angular and spectral ASE transport model. The effective ASE length was set to the pump radius, following TD ASE scaling models in which the relevant ASE gain length is tied to the transverse pumped dimension. This choice should be interpreted as a radius-scale effective path length for an uncapped, non-ASE-suppressed disk, rather than as a resolved angular ASE transport length\cite{ASE3, ASE5}. Since the ASE spectrum, angular distribution, and escape path were not independently measured, $\boldsymbol{\Phi}_\mathrm{ASE}$ should be interpreted as an effective photon flux used to assess the sensitivity of inversion and heat load to ASE-induced stimulated emission. Let $I_0(\cdot)$ and $I_1(\cdot)$ denote the modified Bessel functions of the first kind. The ASE term follows the effective-length treatment commonly used for uncapped Yb:YAG TD gain media\cite{Peterson2011,ASE2}.

\begin{align}\label{PhFlux_ASE}
\begin{split}
\boldsymbol{\Phi}_\mathrm{ASE} = \dfrac{N_2}{\tau _f}L_\mathrm{ASE}\cdot\exp{(g_k\cdot L_\mathrm{ASE}/2)}
\\  \cdot\big(I_0(g_k\cdot L_\mathrm{ASE}/2) -I_1(g_k\cdot L_\mathrm{ASE}/2)\big)
\end{split}
\end{align}

\subsection{Intracavity signal train}
\label{ICSig}

The intracavity signal calculation is introduced to determine the signal photon flux that interacts with the thin disk and, after output coupling, gives the laser output power. To avoid ambiguity, three signal-related quantities are distinguished in the model. The array $\boldsymbol{\Phi}_\mathrm{sig}(t_k)$ represents the traveling-wave photon-flux distribution inside the resonator over one discretized round trip. The scalar $\boldsymbol{\Phi}_\mathrm{l}(t_k)$ denotes the local signal-wavelength photon flux interacting with the thin disk and is used in the rate equation to update the population inversion and stimulated emission. Finally, $P_\mathrm{sig}(t_k)$ is the output-coupled signal power, obtained from the photon flux at the output-coupler position ($j_\mathrm{OC}$) after applying the output-coupler transmission. Therefore, the calculation proceeds by propagating $\boldsymbol{\Phi}_\mathrm{sig}(t_k)$ through the resonator, evaluating $\boldsymbol{\Phi}_\mathrm{l}(t_k)$ at the disk position, updating the gain and absorption through the rate equation, and extracting $P\mathrm{sig}(t_k)$ after convergence.

The intracavity photon flux is represented by a discrete traveling-wave array to obtain a steady-state solution for resonator-mode extraction. Time discretization is required to resolve absorption and stimulated-emission dynamics during multi-pass pump and to define round-trip loss consistently. Without an explicit time grid, the per-pass evolution of absorption, gain saturation, and ASE inside the multi-pass sequence cannot be evaluated consistently. We divide one cavity round trip into $N$ propagation array with $L_c$ and define the intracavity traveling-wave photon-flux array. 

\begin{align}\label{PhFlux_Sig_Init}
\begin{split}
\boldsymbol{\Phi}_\mathrm{sig}(t_k) \equiv\boldsymbol{\Phi}_{k}^\mathrm{sig}= \begin{bmatrix} \Phi_{\mathrm{1},k}\ \Phi_{\mathrm{2},k}\; ...\; \Phi_{\mathrm{N},k}\end{bmatrix}^\mathrm{T},\quad\Phi_\mathrm{j,k}\in\mathbb{R}
\end{split}
\end{align}

The signal train is advanced sequentially through gain, reflection, loss, and output-coupling elements. The detailed numerical implementation of this propagation is provided in Appendix \ref{AppA_IC_Train}. Because the TD is operated in reflection, the resonator signal interacts with the disk over two consecutive propagation arrays around the disk position. The photon flux that enters the rate equation is therefore not the full intracavity array, but the local disk-interaction flux position $j_g$ and $j_g+1$ due to reflection. Including the effective ASE photon flux, the local signal-wavelength photon flux ($\boldsymbol{\Phi}_\mathrm{l}(t_k)$) is written as

\begin{align}\label{PhFlux_Sig}
\begin{split}
\boldsymbol{\Phi}_\mathrm{l}(t_k) = \boldsymbol{\Phi}_{j_g,k} + \boldsymbol{\Phi}_{j_g+1,k} + \boldsymbol{\Phi}_\mathrm{ASE}
\end{split}
\end{align}

where $\boldsymbol{\Phi}_\mathrm{ASE}$ represents the effective ASE photon flux, and this quantity is distinct from the photon flux reaching the output-coupler. The signal output power ($P_\mathrm{sig}$) is calculated using the output-coupled photon flux with the output-coupler transmission $T_\mathrm{OC}$.

\begin{align}\label{SigOutP}
\begin{split}
P_\mathrm{sig}(t_k) = T_\mathrm{OC} \Phi_{j_\mathrm{OC},k} A_\mathrm{l}h\nu_\mathrm{L}\quad \text{where}\quad A_\mathrm{l} = A_\mathrm{p}
\end{split}
\end{align}

The $j_\mathrm{OC}$ is the position of the array where output-coupler is located, and $\Phi_{j_\mathrm{OC},k}$ is the signal photon flux before entering the output-coupler. After each update, the population inversion, gain, and pump absorption coefficient are recalculated. 

\begin{align}\label{BetaUpt_AP1}
\begin{split}
\beta(t_k) = \beta(t_{k-1})+\Delta\beta
\end{split}
\end{align}

The iteration is terminated when the local disk-interaction photon flux satisfies

\begin{align}\label{Converge}
\begin{split}
\dfrac{|\boldsymbol{\Phi}_\mathrm{l}(t_k)-\boldsymbol{\Phi}_\mathrm{l}(t_{k-1})|}{\boldsymbol{\Phi}_\mathrm{l}(t_{k-1})} < 10^{-3}\quad \implies t_k = t_{ss}
\end{split}
\end{align}

The iteration is stopped when it reaches the steady-state time ($t_{ss}$). Only the converged steady-state values are used for comparison with residual pump and output-power measurements. The typical example of simulation results are shown in figure \ref{figA1} in \ref{AppA_IC_Train} for absorption coefficient, residual power, population inversion and single pass gain in time domain, changing from NL regime to MM operation.

\subsection{Volumetric heat load}
Population inversion, absorption, and gain are updated according to the discrete time ($t_k$). Accordingly, the volumetric heat load generated by net energy transition can be derived as Eq. (\ref{HeatLoad}). 

\begin{align}\label{HeatLoad}
\begin{split}
\dot{Q} = -h\nu_\mathrm{L}\gamma_{12}N_2+h\nu_\mathrm{P}(\sigma_\mathrm{a}^\mathrm{(p)}\cdot N_1-\sigma_\mathrm{e}^\mathrm{(p)}\cdot N_2)\boldsymbol{\Phi}_\mathrm{p}(t_\mathrm{ss})\\ +h\nu_\mathrm{L} (\sigma_\mathrm{a}^\mathrm{(l)}\cdot N_1-\sigma_\mathrm{e}^\mathrm{(l)}\cdot N_2)\boldsymbol{\Phi}_\mathrm{l}(t_\mathrm{ss})
\end{split}
\end{align}

Here, $\dot{Q}$ denotes the volumetric heat load generated in the gain medium. The second term describes the net pump-band energy exchange, including pump absorption and pump-band stimulated emission. The third term describes the net signal-band energy exchange, including signal-wavelength reabsorption, stimulated emission into the resonator mode, and, when included, ASE-induced stimulated emission. Therefore, the value of $\dot{Q}$ depends on how the signal-band photon flux $\boldsymbol{\Phi}_\mathrm{l}(t_{ss})$ is defined in the heat-load calculation.

In order to evaluate the volumetric heat load, we use two limiting definitions of $\Phi_\mathrm{l}(t_{ss})$ to estimate the uncertainty associated with ASE escape. When only the coherent resonator signal is included ($\boldsymbol{\Phi}_\mathrm{l} = \boldsymbol{\Phi}_{j_g,k} + \boldsymbol{\Phi}_{j_g+1,k}$) in calculating the heat load $(\dot{Q})$, the heat-load calculation assumes that ASE generated inside the disk does not leave the gain volume as an effective energy-removal channel. This case corresponds to local reabsorption, radiation trapping, or eventual conversion of ASE-related energy into heat, and therefore gives an upper estimate of the deposited heat. 

In contrast, when the effective ASE photon flux is added ($\boldsymbol{\Phi}_\mathrm{l} = \boldsymbol{\Phi}_{j_g,k} + \boldsymbol{\Phi}_{j_g+1,k} + \boldsymbol{\Phi}_\mathrm{ASE}$), the ASE photons are treated as an additional signal-band extraction channel. If these photons escape from the disk without reabsorption, part of the stored energy leaves the local gain volume radiatively, and the calculated heat load gives a lower estimate. The actual deposited heat is expected to lie between these two limits, depending on ASE escape, fluorescence escape, reabsorption, radiation trapping, non-radiative relaxation, and surface- or coating-localized absorption. In practice, the ASE escape fraction can depend on the disk side geometry, surface shape, coating design, and the presence or absence of ASE-suppression structures~\cite{ASE2}. Direct validation of the absolute heat load by temperature, deformation, or wavefront measurement is outside the scope of the present work.

\section{Numerical implementation}

The derived equations for predicting pump absorption and laser performance under experimentally relevant conditions were numerically implemented (MATLAB, 2025a). The numerical implementation explicitly accounted for the pump incidence angle, the spectral property of the pump diode, and the temperature of the thin-disk. Absorption and emission cross-section values measured in \cite{MeasYBYAG} are used along with the temperature dependencies for the numerical calculations. The pump spectrum was centered near 969$\mathrm{nm}$, where the signal was 1030$\mathrm{nm}$. The signal linewidth was treated as an effective spectral width of 0.4 $\mathrm{nm}$ full width at half maximum (FWHM), for evaluating the temperature-dependent cross sections. Since the signal and ASE spectra were not independently measured, the ASE-related contribution should be interpreted as a model sensitivity rather than as a directly measured ASE power. 

The total round-trip loss $\alpha_\mathrm{c}$ is 0.1 $\%$ as well as the reflection loss ($\alpha_\mathrm{HR}$). The output-coupler transmission ($T_\mathrm{OC}$) was set to 5$\%$. The reflectivity of the parabolic mirror and deflection prism in the thin-disk module ($R_\mathrm{PM})$ was assumed to be 99.9$\%$. The pump spot size was 2.8 mm, and the theoretically derived eigenmode size was 0.6 mm, which is significantly smaller than the pump spot size. Thus, the signal would be in multimode regime, and the signal area in the rate equation was treated as an effective multimode overlap area comparable to the pumped area, shown in figure \ref{figAppC}(c) appendix \ref{APPB_Param}. 

The disk temperature was measured (Teledyne, FLIR E50), and the corresponding averaged value was applied in the temperature-dependent formula for NL and MM shown in Table \ref{Tab_Param}, appendix \ref{APPB_Param}. The parameters used in the calculation are described in Appendix \ref{APPB_Param}. The spectral property of pump diode, measured pump profile of NL and MM is shown in figure \ref{figAppC} appendix \ref{APPB_Param}.

\section{Experimental setup}
The experimental setup was designed using a thin-disk module and simple laser system to compare the NL condition and MM operation. The setup as shown in figure \ref{fig1}(c) is capable of simultaneously measuring residual pump power and signal output power, and each power meter is calibrated before the experiment. The thin-disk module was configured with 32 pump-passes. Two different Yb:YAG disks were employed to verify the accuracy of the model: TD1 had a thickness of 189-$\mathrm{\mu m}$ thickness, 8.0 $\mathrm{at.\,} \%$ doping and TD2 as 215-$\mathrm{\mu m}$ thickness, 7.0 $\mathrm{at.\,}\%$ doping. The resonator used is a V-shaped cavity, and the distance between the mirror, the TD, and the output-coupler is 150 mm in each case. The residual pump power measurement is spatially integrated. Therefore, the comparison between the experimental result and calculation validates the integrated absorption budget after the multi-pass sequence, but not the transverse residual-pump profile or pass-resolved spatial redistribution. By utilizing the measured residual pump power, the signal output power can be predicted more accurately, and the amount of heat generated can be calculated indirectly.

\begin{figure*}[!ht]
\centering\includegraphics[width=\textwidth]{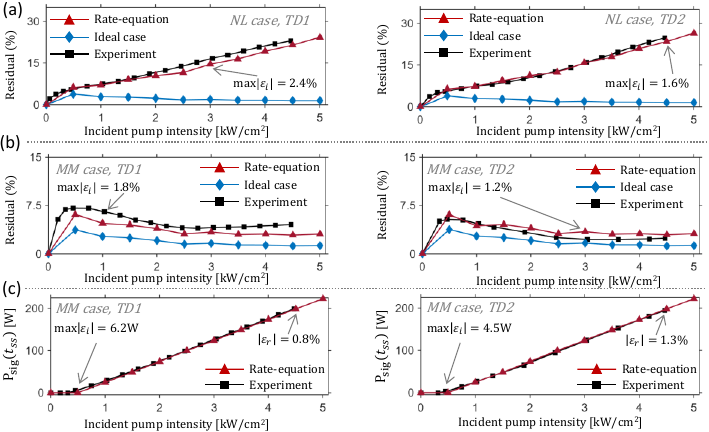}
\caption{\label{fig2} The analysis of the result for TD1 and TD2, in non-lasing (NL) and multimode lasing (MM) case. The residual-pump fraction is defined as $P_\mathrm{res}/P_0$. (a) Residual-pump fraction in the NL case. (b) Residual-pump fraction in the MM case. (c) Measured and quasi-three-level rate-equation based signal output power $P_\mathrm{sig}$  under MM condition. The output-power deviation is reported as absolute power error and relative error over the high-power operating range. The ideal case considers only absorption without taking population inversion into account at room temperature ($\alpha_\mathrm{eff} = \alpha_0 = N_0 \sigma_\mathrm{a}^\mathrm{(p)}$). The $\mathrm{max}|\varepsilon_i|$ stands for maximum absolute error, difference between simulation value and experimental value. $|\varepsilon_r|$ is absolute relative error. } 
\end{figure*}

\begin{figure}[!hb]
\centering
\includegraphics[width=0.98\columnwidth]{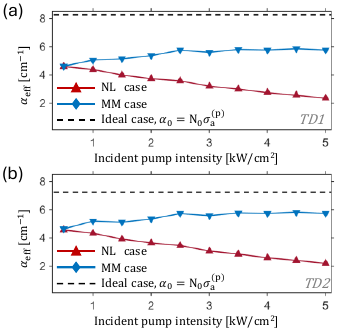}
\caption{\label{fig3} Effective pump absorption coefficient evaluated under NL and MM operation of (a) TD1 and (b) TD2. The dashed line denotes the room-temperature small-signal absorption coefficient, $\alpha_0 = N_0\sigma_\mathrm{a}^\mathrm{(p)}$. The difference between the calculated operating-state absorption and the small-signal value arises from disk temperature, pump-band emission, population inversion, resonator-mode extraction, and ASE-induced inversion depletion. } 
\end{figure}

\section{Results and discussion}
\subsection{Residual-pump evaluation of the absorption budget}

The residual pump fraction provides the primary experimental check of the absorption budget in the multipass pump module. Figure \ref{fig2} compares the measured and calculated residual pump fractions for the two disks under NL and MM operation. The agreement indicates that the pass-integrated pump absorption is captured over the investigated pump-intensity range. The output-coupled signal power is then used as an additional consistency check, confirming that the same absorption state also describes resonator-mode extraction in the high-power operating range.

The calculated residual pump fraction agrees with the measurement within 2.4$\%$ in NL and 1.8$\%$ in MM operation, with the largest deviation at low pump power as shown in figure \ref{fig2}(a) and (b). The rate-equation based output power calculation also agrees with the measurement, with a relative error ($|\varepsilon_r|$) of within 1.3$\%$, except for the low pump intensity, shown in figure \ref{fig2}(c). The error at low pump intensity is expected because the residual is strongly governed by the absorption estimate with the signal and ASE spectral assumptions. Here, the signal and ASE spectral widths were fixed to 0.4 nm, which can increase the error in the residual, whereas the output-power error is partially suppressed by the output-coupler transmission. The remaining mismatch is therefore attributed mainly to uncertainties of TD coatings, spatial uniformity of the beam, and the spectral property of ASE. Nevertheless, the theoretical approach showed substantial agreement with the results, supporting the use of selecting the ASE length as the pump radius, as reported in previous studies\cite{Peterson2011,Antognini2009,ASE2}. The defined 0.1$\%$ cavity loss will primarily consist of reflection and scattering losses. 

Figure \ref{fig3} is the main diagnostic result of the residual-pump analysis. It shows that the effective pump absorption coefficient under operation differs from both the non-lasing state and the room-temperature small-signal coefficient. In practice, multi-pass pump module are designed considering the medium’s ideal absorption coefficient. However, as can be inferred from figure \ref{fig3}, the effective absorption coefficient derived when the laser is actually operating differs from the ideal absorption coefficient. It turns out that this is due to the medium’s temperature, reabsorption in the 969 $\mathrm{nm}$ pump band, and population inversion during lasing. For the present TDs and pump geometry, this difference corresponds to an absorption-budget allowance of approximately 10–20$\%$ when estimating residual pump or selecting the number of pump-passes. This value should not be interpreted as a universal design margin. Nevertheless, it demonstrates that pump-pass design based only on room-temperature small-signal absorption can be insufficient for high-power operation.

\begin{figure*}[t]
\centering\includegraphics[width=2\columnwidth]{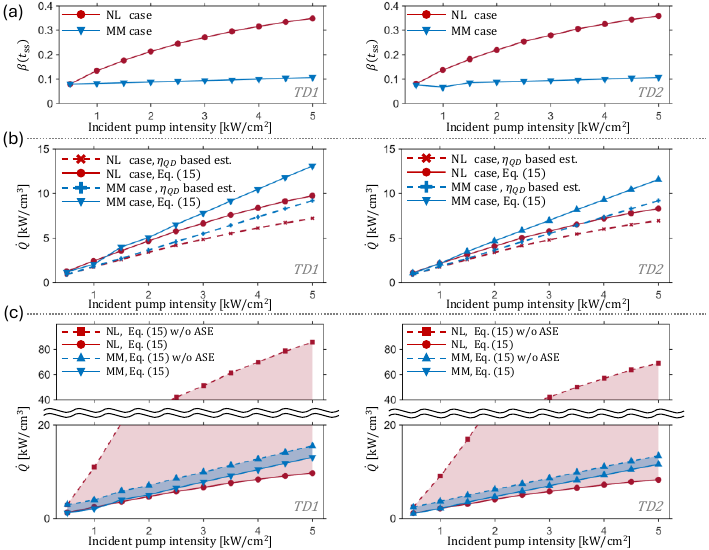}
\caption{\label{fig4} Steady-state population inversion ratio ($\beta(t_\mathrm{ss})$) and volumetric heat load $\dot{Q}$ in steady-state for TD1 and TD2 respectively. (a) Population inversion in the steady-state time in NL and MM case. (b) Comparison of volumetric heat load between quantum defect-based and from Eq. (\ref{HeatLoad}) in NL and MM case. The quantum-defect $(\eta_\mathrm{QD}=1-\lambda_\mathrm{p}/\lambda_\mathrm{l})$ is the applied to the absorbed pump power. (c) Volumetric heat load with ASE and without ASE in NL and MM case. Note that the volumetric heat load calculated without ASE means that all generated ASE is absorbed and converted into heat.} 
\end{figure*}

\subsection{Heat-load implications of residual-pump}

The calculation includes ASE as an effective inversion-depletion channel~\cite{ASE1,ASE2,ASE3,ASE4}. Because the ASE spectrum, angular distribution, and escape path were not independently measured, the ASE contribution should be interpreted as a model-based sensitivity rather than as a direct measurement of ASE power. Within this treatment, ASE influences the steady-state inversion and therefore the residual-pump-based absorption estimate, whereas resonator-mode extraction dominates the inversion depletion above threshold.

Figure \ref{fig4} presents the calculated heat load ($\dot{Q}$) at steady-state,  computed from Eq. (\ref{HeatLoad}) with the population inversion ratio $\beta(t_{ss})$ in the NL and MM case. The predicted result is consistent with values reported in the literature~\cite{Peterson2011}, lending credibility to the underlying energy-balance assumptions. As shown in figure \ref{fig4} (a), population inversion in the steady state remains higher under NL condition. The calculated heat load increases with pump power and can be decomposed into contributions from pump absorption, signal extraction, and ASE. The heat load calculated from Eq. (\ref{HeatLoad}) is higher than quantum-defect based estimation, because it includes population-dependent pump absorption, laser-wavelength reabsorption, stimulated-emission balance, and ASE-related energy exchange.

In the NL case, Eq. (\ref{HeatLoad}) can give a lower estimate of the deposited heat if ASE or fluorescence is reabsorbed inside the disk, because there is no well-defined resonator mode extracting the radiative energy. As shown in figure \ref{fig4} (b) and (c), the heat loads calculated with and without ASE define a range for the deposited heat, depending on how much ASE and fluorescence escape or are reabsorbed. This distinction is particularly important in the NL case, where ASE and fluorescence are not extracted through a well-defined resonator mode and can be partially trapped or re-absorbed in the gain medium. In contrast, under MM condition, the resonator mode provides a well-defined stimulated-emission extraction channel, making Eq.  (\ref{HeatLoad}) more directly applicable for evaluating the local energy exchange associated with lasing.

Under the adopted energy-balance assumptions, the heat-load estimate from Eq. (15) is up to approximately 40\% higher than the simple quantum-defect estimate in MM operation, shown in figure \ref{fig4}(b). In addition, the heat loads calculated with and without ASE are similar in the MM case in figure \ref{fig4}(c). In the present work, the TDs used here were not designed with a structure that allows ASE photons to escape effectively. By improving ASE escape, the signal output can be enhanced\cite{ASE1,ASE2,ASE4}. 
The comparison with the surface-temperature estimate from Eq. (\ref{1DTemp}) suggests that part of the ASE-related radiative energy may be reabsorbed and deposited as heat, shown in figure \ref{figAppD} in \ref{APPB_Param}. The agreement between measured and calculated residual pump fractions supports the use of a pump-radius-scale $L_\mathrm{ASE}$ in the present uncapped disks. 

\begin{figure*}[!t]
\centering\includegraphics[width=0.95\textwidth]{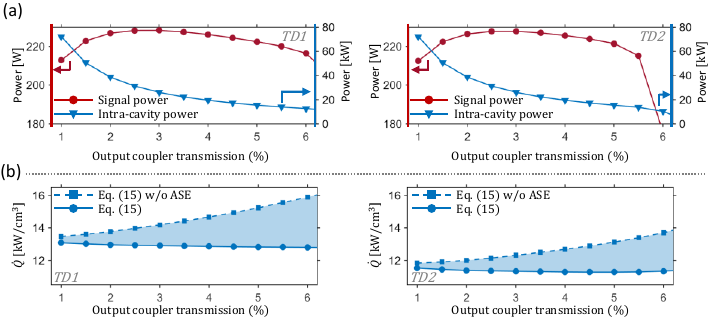}
\caption{\label{fig5} Output coupler transmission dependence in the incident pump intensity 4.5 $\mathrm{kW/cm^2}$. (a) Signal power, intracavity signal power for TD1 and TD2 in the MM case. (b) Calculated volumetric heat load ($\dot{Q}$) with ASE and without ASE for TD1 and TD2 in the MM case. Note that the volumetric heat load calculated without ASE means that all generated ASE is absorbed and converted into heat.} 
\end{figure*}

\begin{figure*}[!h]
\centering\includegraphics[width=0.95\textwidth]{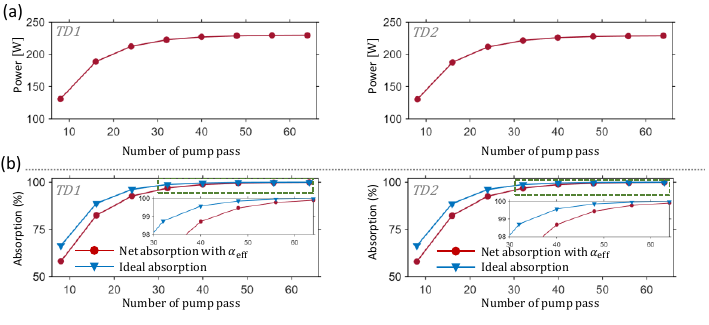}
\caption{\label{fig6} Pump-pass dependence in the incident pump intensity 4.5 $\mathrm{kW/cm^2}$. (a) Calculated signal power for TD1 and TD2 in the MM case. (b) Net absorption with $\alpha_\mathrm{eff}$ and ideal absorption ($\alpha_\mathrm{eff} = \alpha_0 = N_0\sigma_\mathrm{a}^\mathrm{(p)}$) for TD1 and TD2 in the MM case.} 
\end{figure*}
 
\subsection{Design trends for output coupling and pump-pass number}

We analyze the design parameter dependence and it is not intended to provide an experimentally verified global optimum for $T_\mathrm{OC}$ or pump-pass number. The TD temperature, mode overlap, cavity loss, and ASE treatment are held fixed, whereas in a real high-power system these quantities can change with output coupling, thermal deformation, and multimode redistribution. Therefore, the results should be interpreted as tendency showing trade-offs among output power, intracavity power, residual pump absorption, and heat load.

Figure \ref{fig5} analyzes the signal output from the laser system, the signal power inside the cavity, and the heat load as a function of the output-coupler transmission. In this analysis, the measured temperature is held constant, and the incident pump intensity is 4.5 $\mathrm{kW/cm^2}$. In the present configuration, the calculated output power shows a broad optimum at low-to-moderate transmission, while the intracavity power decreases as $T_\mathrm{OC}$ increases, shown in figure \ref{fig5}(a). Therefore, selecting $T_\mathrm{OC}$ solely from output power can be misleading; the associated intracavity power and heat load should also be considered, as shown in figure \ref{fig5}(b). While figure \ref{fig5} suggests that a lower transmission rate for a given output coupler can achieve maximum output, it is important to consider that the intracavity signal power is much higher than that outside the resonator, which can more easily lead to damage to other optical components or coatings. 

In addition, the pump-pass should be implemented using optimal parameter settings based on the in-situ absorption coefficient. In laser oscillation, a signal is generated due to stimulated emission, and as a result, the population inversion ratio remains low in the steady state, and accordingly, the effective absorption rate increases, shown in figure \ref{fig3}. The calculated output power and net absorption approach an asymptote as the pump-pass number increases, shown in figure \ref{fig6}(a). For the present disks, pump geometry and 5$\%$ output-coupler transmission, more than 50 passes are required to approach the high-absorption regime($>99\%$), as shown in figure \ref{fig6}(b). In practice, the optimum pass number should balance residual pump reduction against additional optical loss, alignment sensitivity, module complexity, and changes in the spatial heat-source distribution. Although the temperature of the medium would vary depending on the output-coupler transmission and the pump-pass number, it is sufficient for identifying the general trend.

\section{Conclusion}
This work demonstrates how residual pump power can be used as a diagnostic measurement for absorption in multi-pass pumped Yb:YAG thin-disk lasers. Rather than treating the absorption coefficient as a fixed material parameter, the presented formulation links residual pump, population inversion, resonator-mode extraction, and heat load. The analysis shows that multimode lasing can increase the absorption by depleting the inversion. It also shows that quantum-defect based heat load may underestimate the volumetric heat load because it neglects population-dependent absorption, reabsorption, and ASE-induced transitions. The calculated heat load should not be interpreted as a validated temperature prediction; absolute thermal validation would require spatially resolved optical fields, ASE/fluorescence transport, coating absorption, and thermo-mechanical boundary conditions. Within these limits, the residual pump based approach provides practical guidance for pump-module design, output coupling, and ASE management in high-power thin-disk laser scaling.

\section{Funding}
This research was co-funded by the European Union (MERIT - Grant Agreement No. 101081195) and European Union and the state budget of the Czech republic under the project LasApp CZ.02.01.01/00/22$\_$008/0004573.

\section{Disclosures}
The authors declare no conflicts of interest.

\section{Data Availability Statement}
The data that support the findings of this study are openly available in Zenodo.

\appendix
\section{Intracavity photon flux}
\label{AppA_IC_Train}

\begin{figure*}[b]
\centering\includegraphics[width=\textwidth]{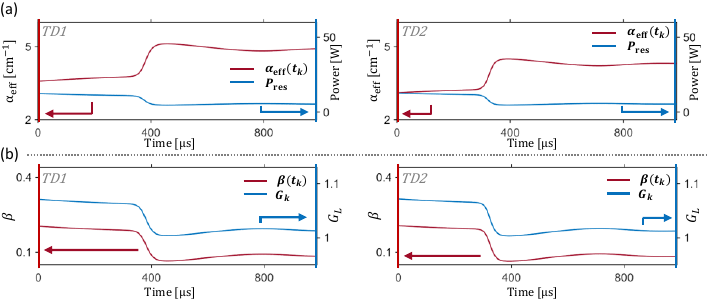}
\renewcommand{\thefigure}{A\arabic{figure}}
\setcounter{figure}{0}
\caption{\label{figA1} Typical example of time-domain simulation of population-inversion dynamics in MM operation when incident pump intensity is 3 $\mathrm{kW/cm^2}$. (a) The effective absorption coefficient ($\alpha_\mathrm{eff}$), residual pump power after multi-pass pumping ($P_\mathrm{res}$) in temporal domain for thin-disk 1 (TD1) and (b) thin-disk 2 (TD2) respectively. (b) The population inversion ($\beta$) and single-pass gain ($G_k$) of the TD1 and TD2, respectively.} 
\end{figure*}

The intracavity signal is represented by a discretized traveling photon-flux train to obtain a steady-state solution for resonator-mode extraction. The discretization is used as a numerical relaxation scheme rather than as a measured temporal waveform. One cavity round trip is divided into $N$ propagation arrays,

\renewcommand{\theequation}{A\arabic{equation}}
\setcounter{equation}{0}
\begin{align}\label{PhFlux_Sig_Init_AP}
\begin{split}
\Phi_\mathrm{sig}(t_k) \equiv\Phi_{k}^\mathrm{(sig)}= \begin{bmatrix} \Phi_{\mathrm{1},k}\ \Phi_{\mathrm{2},k}\; ...\; \Phi_{\mathrm{N},k}\end{bmatrix}^\mathrm{T},\\ \Phi_\mathrm{j,k}\in\mathbb{R}
\end{split}
\end{align}

where $\Phi_\mathrm{sig}(t_k)$ is the intracavity traveling-wave signal represented by an $N$-th array representing one complete round trip sampled at discrete time $t_k$. The $j$-th entry of the photon-flux vector denotes the signal photon flux contained in the $j$-th spatial bin at time $t_k$, and $\Phi_\mathrm{j,k}$ denotes photon flux contained in the $j$-th array at time $t_k$. Advancing from $t_k$ to $t_{k+1}$ corresponds to propagation by one discretized spatial step, implemented as a cyclic shift. Propagation by one time step is represented by permutation matrix $\mathbf{S}$ as a cyclic right shift $\Phi\leftarrow\mathbf{S}\Phi$ that,

\begin{align}\label{ShiftOpt_AP}
\begin{split}
(\mathbf{S}\mathrm{x})_1 = x_N,\quad(\mathbf{S}\mathrm{x})_j = x_{j-1}\quad(j=2,...,N)
\end{split}
\end{align}

Inside the cavity, the signal photon flux propagates and reverses its propagation direction upon reflection at the cavity boundaries. We therefore distinguish between the forward-propagating photon-flux array, denoted by $\Phi_{k}^{\mathrm{sig}(+)}$, and the reverse-propagating photon-flux array, denoted by $\Phi_{k}^{\mathrm{sig}(-)}$. The reverse ordering of the discretized arrays is represented by the anti-diagonal identity matrix $\mathrm{J}$, such that index reversal is written as Eq. (\ref{SigPhotonRF_AP}) when switching propagation direction.

\begin{align}\label{SigPhotonRF_AP}
\begin{split}
\Phi_{k}^{\mathrm{sig}(-)} \leftarrow \mathrm{J}\Phi_{k}^{\mathrm{sig}(+)},\quad 
\Phi_{k}^{\mathrm{sig}(+)} \leftarrow \mathrm{J}\Phi_{k}^{\mathrm{sig}(-)}
\end{split}
\end{align} 

The thin-disk single-pass gain $G_k = \exp{(g_k\cdot L_\mathrm{d})}$ from Eq. (\ref{SSGain}) occupies two adjacent array elements in the diagonal mask. The index $j_\mathrm{g}$ denotes the bin occupied by the thin-disk gain medium. The adjacent bin $j_\mathrm{g}+1$ is also included to account for the reflected interaction at the same gain location in forward diagonal matrix $D_k^{(+)}$. The index $j_\mathrm{OC}$ denotes the bin where output-coupler is located . For reverse diagonal matrix, spatial bin occupied at $N-j_\mathrm{g}, N+1-j_\mathrm{g}, N+1-j_\mathrm{OC}$ in $D_k^{(-)}$. 

\begin{align}\label{Mask_AP}
\begin{split}
D_k^{(+)}=\text{diag}\big(1,...,G_k, G_k, ..., 1, 1-\alpha_\mathrm{HR} \big),\\ D_k^{(-)}=\text{diag}\big(1,...,G_k, G_k, ..., 1, 1-\alpha_\mathrm{OC} \big)
\end{split}
\end{align}

with the boundary reflectivity placed at the last array  at forward $1- \alpha_\mathrm{HR} = 0.999$, and reverse $\alpha_{OC} = T_\mathrm{OC}+\alpha_\mathrm{c}$ with output-coupler transmission $T_\mathrm{OC}$, and cavity loss $\alpha_\mathrm{c}$. The diagonal structure reflects the fact that gain and boundary losses act locally on the photon flux already contained in each spatial bin, without introducing coupling between different arrays. Consequently, all the arrays outside the thin-disk position and the cavity boundaries remain unchanged during this sub-step, and their diagonal entries are equal to 1. The spontaneous emission injected into the two gain array is modeled by 

\begin{align}\label{PhFlux_SE_AP}
\begin{split}
\mathrm{b}_k^{(+)}  =\Delta\Phi_\mathrm{SE}(\mathrm{e}_{j_g}+\mathrm{e}_{j_g+1})
\end{split}
\end{align}

Here, $e_j\in\mathbb{R}^N$ denotes the $j$-th canonical basis vector, whose $j$-th entry is unity and all other entries are zero. The reverse is defined by anti-diagonal identity matrix $\mathrm{J}$ as $b_k^{(-)} = \mathrm{J}b_k^{(+)}$. Then the update equations are simply 

\begin{align}\label{PhFlux_SigUpt_AP}
\begin{split}
\Phi_{k+1}^{(+)} = D_k^{(+)}\mathbf{S}\Phi_k^{(+)}+b_k^{(+)},
\\ \Phi_{k+1}^{(-)} = D_k^{(-)}\mathbf{S}\Phi_k^{(-)}+b_k^{(-)}
\end{split}
\end{align}

The thin-disk gain medium is located at $j_g$ and $j_g+1$ due to reflection (i.e. the signal photon flux is overlapped due to reflection), then the population inversion ratio $\beta(t_k)$ is updated by Eq. (\ref{RateEq}).

\begin{align}\label{PhFlux_Sig_AP}
\begin{split}
\Phi_\mathrm{l}(t_k) = \Phi_{j_g,k} + \Phi_{j_g+1,k} + \Phi_\mathrm{ASE}
\end{split}
\end{align}

\begin{align}\label{BetaUpt_AP}
\begin{split}
\beta(t_k) = \beta(t_{k-1})+\Delta\beta
\end{split}
\end{align}

In this case, due to the updated population inversion, the absorption coefficient and single-pass gain are re-evaluated. The signal output power is then calculated as 

\begin{align}\label{SigOutP_AP}
\begin{split}
P_\mathrm{sig}(t_k) = T_\mathrm{OC} \Phi_{j_\mathrm{OC},k} A_\mathrm{l}h\nu_\mathrm{L}
\end{split}
\end{align}

The iteration of numerical calculation terminates when the signal photon flux $(\Phi_\mathrm{l})$ converges. The numerical implementation for population inversion, absorption and gain are then updated and terminated when it reaches the steady-state $t_\mathrm{ss}$.  

The temporal evolution of absorption, residual pump power, population inversion, and gain differs fundamentally between NL and MM operation because the intracavity signal introduces strong stimulated-emission feedback, as shown in figure \ref{figA1} (a) and (b). In the early stage of the time domain, population inversion remains high and it is indicated as NL operation. Under NL condition, a substantial fraction of the incident pump remains unabsorbed, so the residual power stays high and the dominant additional loss channel is attributed to ASE rather than stimulated extraction in early time domain. In contrast, under MM condition the intracavity signal photon flux rapidly rises by accelerating inversion depletion and drives a concurrent increase in effective absorption. The gain then quickly approaches a steady-state when it is equal to the total cavity loss including transmission of OC and cavity loss, known as gain clamping, and the inversion correspondingly converges to the steady-state as well. This behavior results in faster convergence in the time domain at high pump intensity due to higher spontaneous emission and gain values. 

\begin{align}\label{Converge_AP}
\begin{split}
\dfrac{|\Phi_\mathrm{l}(t_k)-\Phi_\mathrm{l}(t_{k-1})|}{\Phi_\mathrm{l}(t_{k-1})} < 10^{-3}
\end{split}
\end{align}

\begin{figure}[!b]
\renewcommand{\thefigure}{B\arabic{figure}}
\setcounter{figure}{0}
\centering\includegraphics[width=0.97\columnwidth]{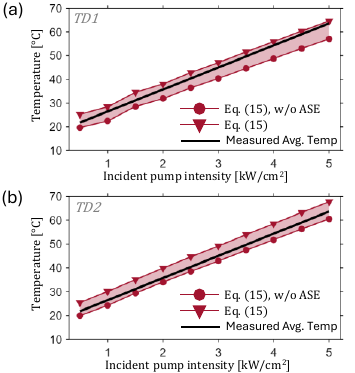}
\caption{\label{figAppD} Calculated surface temperature of (a) TD1 and (b) TD2, respectively. The measured average temperature(black line) is located in between calculated heat load with ASE and without ASE.} 
\end{figure}

\section{Parameters}
\label{APPB_Param}

Note that the absorption and emission cross-section data listed in Table. \ref{Tab_Param} are typical measured values based on the assumptions of specific temperature and ideal monochromatic radiation. In actual simulations, the absorption and emission cross-section data were derived for the corresponding temperatures and spectra.

\begin{table*}[!htbp]
\scriptsize
\renewcommand{\thetable}{B\arabic{table}}
\centering
\caption{Simulation parameters}
\label{Tab_Param}
\begin{tabular}{@{}lcccc@{}}
\hline
Symbol & unit & value & description & Notes \\
\hline
$\sigma_\mathrm{a}^\mathrm{(p)}$ & $\mathrm{cm^2}$ &\makecell{7.54$\times 10^{-21}$ at 27$^\circ$C\\5.11$\times 10^{-21}$ at 100$^\circ$C}&Absorption cross-section at 969 nm& measured value used \cite{MeasYBYAG} \\
$\sigma_\mathrm{a}^\mathrm{(l)}$ & $\mathrm{cm^2}$ &\makecell{1.25$\times 10^{-21}$ at 27$^\circ$C\\1.53$\times 10^{-21}$ at 100$^\circ$C}& Absorption cross-section at 1030 nm& measured value used \cite{MeasYBYAG} \\
$\sigma_\mathrm{e}^\mathrm{(p)}$ & $\mathrm{cm^2}$ &\makecell{6.88$\times 10^{-21}$ at 27$^\circ$C\\4.66$\times 10^{-21}$ at 100$^\circ$C}& Emission cross-section at 969 nm& measured value used \cite{MeasYBYAG} \\
$\sigma_\mathrm{e}^\mathrm{(l)}$ & $\mathrm{cm^2}$ &\makecell{2.15$\times 10^{-20}$ at 27$^\circ$C\\1.49$\times 10^{-20}$ at 100$^\circ$C}& Emission cross-section at 1030 nm& measured value used \cite{MeasYBYAG} \\
$L_\mathrm{d}$ & $\mu$m & \makecell{189 (TD1)\\215 (TD2)} & TD thickness &- \\
$N_{dop}$ & $\mathrm{at}.\%$ & \makecell{8.0 (TD1)\\ 7.0 (TD2)} & TD doping concentration & -\\
$N_0$ & $\mathrm{cm^3}$ & \makecell{1.107$\times10^{21}$ (TD1)\\0.969$\times10^{21}$ (TD2)} &  Total ion density & - \\
$\tau_f$ & $\mathrm{ms}$ & 0.95 & Fluorescence lifetime & - \\
$\gamma_{12}$ & kHz & 1.052 & Decay rate & $\gamma_{12} = 1/\tau_f$ \\
$n_\mathrm{pass}$ & - & 32 & Number of passes for pumping & - \\
$T_\mathrm{NL}$ & K & $10.5\times I_\mathrm{p}\, \mathrm{[kW/cm^2]}+290.60$ & TD surface temperature & in non-lasing case\\
$T_\mathrm{MM}$ & K & $9.29\times I_\mathrm{p}\, \mathrm{[kW/cm^2]}+290.14$ & TD surface temperature & in multimode lasing case\\
$I_\mathrm{p}$ & $\mathrm{kW/cm^2}$ & 0,1,2,3,4,5 & Incident pump intensity & - \\
$\lambda_\mathrm{p}$ & nm & - & Peak of pump wavelength & Spectrum directly utilized, Figure \ref{figAppC}(a) \\
$\nu_\mathrm{p}$ & Hz & - & Pump optical frequency & $\nu_\mathrm{p} = c/\lambda_\mathrm{p}$ \\
$\lambda_\mathrm{L}$ & nm & 1030$\pm$0.2 & Peak of signal wavelength & - \\
$\nu_\mathrm{L}$ & Hz & 2.91$\times 10^{14}$ & Signal optical frequency & $\nu_\mathrm{L} = c/\lambda_\mathrm{L}$ \\
$\alpha_\mathrm{c}$ & $\%$ & 0.1 & Cavity loss & -\\
$\alpha_\mathrm{OC}$ & $\%$ & 5.1 & Total cavity loss & $\alpha_\mathrm{c} +T_\mathrm{OC}$ \\
$T_\mathrm{Oc}$ & $\%$ & 5 & output-coupler transmission & -\\
$R_\mathrm{PM}$ & $\%$ & 99.9 & Total reflectivity in thin-disk module & - \\
$\alpha_\mathrm{HR}$ & $\%$ & 0.1 & End mirror loss & -\\
$R_p$ & mm & 1.394 & Pump radius, HWHM & see figure \ref{figAppC}(b) and (c)\\

\hline
\multicolumn{5}{c}{\makecell{\textit{Note:}Abbreviation follows; OC: output-coupler, TD: thin-disk, RT: room-temperature, HWHM: half width at half maximum, \\FWHM: full width at half maximum.}}
\end{tabular}
\end{table*}

\begin{figure*}[t]
\renewcommand{\thefigure}{B\arabic{figure}}
\centering\includegraphics[width=\textwidth]{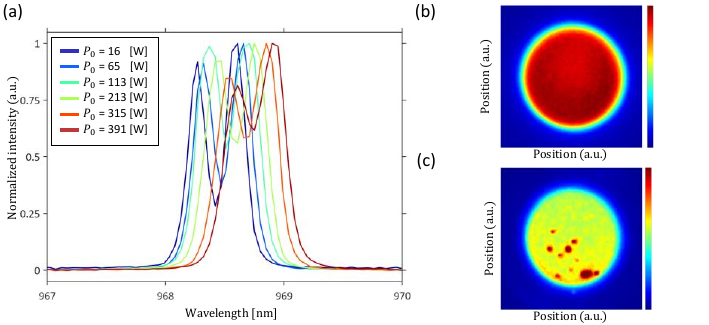}
\caption{\label{figAppC} (a) measured spectrum of laser pump diode. (b) Typical example of measured pump profile in NL case. (c) Typical example of measured pump profile in MM case.} 
\end{figure*}

The reconstructed surface temperature is shown in figure \ref{figAppD}. The result was derived from Eq. (\ref{1DTemp}), which is analytical solution of a one-dimensional steady-state heat-conduction model with uniform volumetric heat generation in the Yb:YAG disk.

\renewcommand{\theequation}{B\arabic{equation}}
\setcounter{equation}{0}
\begin{align}
\label{1DTemp}
\begin{split}
T_\mathrm{Disk}(z) =  \\\frac{\dot{Q}L_\mathrm{d}+h_\mathrm{air}(T_\mathrm{air}-T_\mathrm{c})+(h_\mathrm{air}\dot{Q}L_\mathrm{d}^2)/(2k_\mathrm{YbYAG})}{k_\mathrm{YbYAG} + h_\mathrm{air}L_\mathrm{d}}z - \\ \frac{\dot{Q} z^2}{2k_\mathrm{YbYAG}} + T_\mathrm{c}
\end{split}
\end{align}

The surface temperature corresponds to $T_\mathrm{Disk}(L_\mathrm{d})$). The volumetric heat load ($\dot{Q}$) was taken from the MM case and assumed to be spatially uniform. Thermal radiation and temperature dependence of $k_\mathrm{YbYAG}$ were neglected in this simplified estimate. The $h_\mathrm{air} = 13\mathrm{[W/m^2K]}$, $k_\mathrm{YbYAG}=9.5\mathrm{[W/mK]}$, $T_\mathrm{c}=15^\circ$C for cooling temperature, $T_\mathrm{air}=27^\circ$C for air temperature, natural convection. The Measured spectrum of pump diode and pump profile in NL/MM case is shown in figure \ref{figAppC}.

\clearpage
\bibliographystyle{unsrtnat}
\bibliography{Manu_bib}

\end{document}